\documentclass[pre,showpacs,amssymb,twocolumn]{revtex4}
\usepackage{epsfig}
\usepackage{wasysym}

\newcommand{\2}{{\textstyle \frac{1}{2}}}
\newcommand{\dd}{\mathrm{d}}

\begin{document}

\title{Critical adsorption at chemically structured substrates}
\date{\today}
\author{Monika Sprenger}
\author{Frank Schlesener}
\author{S. Dietrich}
\affiliation{Max-Planck-Institut f\"ur Metallforschung, Heisenbergstr. 3, D-70569 Stuttgart, Germany,}
\affiliation{Institut f\"ur Theoretische und Angewandte Physik, Universit\"at Stuttgart, Pfaffenwaldring 57, D-70569 Stuttgart, Germany}

\begin{abstract}
  We consider binary liquid mixtures near their critical consolute points and exposed to geometrically
  flat but chemically structured substrates. The chemical contrast between the various substrate 
  structures amounts to opposite local preferences for the two species of the binary liquid mixtures.
  Order parameters profiles are calculated for a chemical step, for a single chemical stripe, and for 
  a periodic stripe pattern. The order parameter distributions exhibit frustration across the chemical
  steps which heals upon approaching the bulk. The corresponding spatial variation of the order 
  parameter and its dependence on temperature are governed by universal scaling functions which we
  calculate within mean field theory. These scaling functions also determine the universal behavior
  of the excess adsorption relative to suitably chosen reference systems.
\end{abstract}

\maketitle

\section{\label{sec:intro}Introduction}
%
Chemically structured substrates have gained significant importance within the last years. Since it 
is possible to produce networks of chemical lanes at the $\mu m$-scale and even below these 
chemically structured substrates have applications in micro-reactors, for the ``lab on a chip'', and 
in chemical sensors \cite{Thorsen2002, Juncker2002}. They can operate with small amounts of reactants 
which is important when investigating expensive substances and substances which are available only 
in small amounts like biological material or when dealing with toxic or explosive materials. At 
these small scales the interaction of the fluids with the substrate becomes important and there is 
the challenge to control the distribution and the flow of the fluids on these structures.

\begin{figure}[h]
    \begin{center}
      \epsfig{file=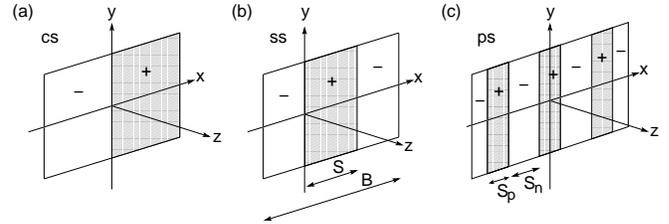, width=0.48\textwidth}
      \caption{\label{fig:geometry}The different chemical substrate structures studied in this article:
               (a) chemical step (cs), (b) single stripe (ss), (c) periodic stripes (ps). All systems
               are translationally invariant in the $y$-direction. The shading indicates different  
               local preferences for, e.g., the two components of a binary liquid mixture exposed 
               to the geometrically flat surface of macroscopicly large lateral extension $B$.}
    \end{center}
  \end{figure}

In the following we investigate three different types of chemically structured substrates, as shown
in Fig.\,\ref{fig:geometry}. First we analyze fluid structures at a chemical step (see Fig. 
\ref{fig:geometry}(a)) which is important for understanding the local properties of fluids at the 
border of stripes. Next we consider a single chemical stripe (see Fig.\,\ref{fig:geometry}(b)) as the 
simplest chemical surface pattern, and finally we study a periodic stripe pattern (see Fig. 
\ref{fig:geometry}(c)) as the paradigmatic case for the investigation of adsorption at heterogeneous 
surfaces.

The chemical contrast on the substrates acts on the adjacent liquid \cite{Boltau1998, Sprenger2003}. 
In order to impress the chemical pattern of the substrate on a one-component liquid, the chemical 
structure has to be chosen as a pattern of lyophobic and lyophilic regions. For binary liquid 
mixtures one chooses a pattern of two different substrate types, such that one component is preferred 
by one substrate type and the second component by the other substrate type.
This lateral structuring of the substrate causes a rich fluid-substrate interface structure which 
typically depends on the molecular details of the local force fields. However, in this study we focus 
on the particular case of the fluid being close to a second-order phase transition. This is either the 
liquid-vapor critical point of a one-component liquid or the critical demixing transition of a binary 
liquid mixture. In these cases the ensuing critical phenomena are to a large extent universal in 
character, i.e., they render molecular details irrelevant in favor of universal scaling functions by 
involving spatial variations on the scale of the diverging correlation length. Near the critical point 
the surface patterning acts like a laterally varying surface field of alternating sign. This generates
an order parameter profile characterizing critical adsorption of opposite sign such that the system is 
frustrated across the chemical steps. Upon approaching the bulk of the fluid this frustration is
healed and the healing is expected to be governed by universal scaling functions. In addition, the 
substrate patterning results in a change of the excess amount of adsorbed fluid with respect to the 
case of critical adsorption on a homogeneous substrate. The excess adsorption is expected to be 
governed by universal scaling functions, too.

It is the purpose of this contribution to describe this scaling in terms of general renormalization 
group arguments and to calculate the corresponding universal scaling functions to lowest order, i.e., 
within mean field theory. If, as it is actually the case, the size of the lateral structures is 
comparable with the range of the correlation length, which can reach up to $100\,nm$ close to the 
critical point, one can expect a rich interplay between the externally imprinted patterns and the 
critical phenomena characterized by the correlation length.

\subsubsection{Critical phenomena}
%
In order to describe critical phenomena \cite{Stanley1971, Zinn-Justin1989} one distinguishes 
properties to classify them. Starting with unconfined systems one introduces the so-called bulk 
universality classes which are characterized by critical exponents and amplitudes which describe the 
dependence of various quantities, e.g., the order parameter and the correlation length, on the 
reduced temperature $t=\frac{T-T_c}{T_c}$ upon approaching the critical temperature $T_c$. These 
critical exponents are universal, i.e., the same for all members of a universality class. The 
amplitudes usually are non-universal, whereas the number of independent non-universal amplitudes is 
limited; any non-universal amplitude can be expressed in terms of these independent non-universal 
amplitudes and universal amplitude ratios.
One-component fluids near their liquid-vapor critical point and binary liquids near their demixing
critical point belong to the Ising universality class like uniaxial ferromagnets. These systems have
two independent non-universal amplitudes. In this sense our subsequent analysis holds for all systems
encompassed by the Ising universality class.

The order parameter $\phi$ indicates the degree of order in the system and has to vanish above the 
critical temperature in the absence of an ordering field while below the critical temperature it takes 
a finite value. It is described by the universal critical exponent $\beta$ and a non-universal 
amplitude $a$: $ \phi(t) = a |t|^{\beta}$. For a one-component fluid $\phi$ is the difference between
the density and its value at the critical point. In the case of a binary liquid mixture it is chosen 
as the difference between the concentration of one fluid component and its concentration at the 
critical demixing point.

The correlation length is defined by the exponential decay of the bulk two-point correlation 
function $G(r)$ for large distances $(r\to \infty$) at temperatures off criticality 
($T\neq T_c$) and it is denoted as $\xi^+$ above and as $\xi^-$ below the critical temperature, 
respectively. It diverges according to the power law $\xi^{\pm} = \xi_0^{\pm} |t|^{-\nu}$ with the 
universal critical exponent $\nu$ and the non-universal amplitudes $\xi_0^{\pm}$. Their ratio 
$\xi_0^+ / \xi_0^-$ is, however, universal. If the critical medium is brought near a substrate, 
surface universality classes come into play which we shall discuss in Sec.\,\ref{sec:hom-sub}.

\subsubsection{Experimental methods}
%
Different techniques to produce chemical structures in the range of $\mu m$ and $nm$ have been
established \cite{Xia1999}. In order to obtain topologically flat but chemically structured substrates 
one can take advantage of the self-ordering mechanism of self-assembled monolayers (SAM) in using 
block-copolymers or mixtures of polymers which form structures while demixing \cite{Bates1991, 
Krausch2002, Boltau1998}. The morphology and domain size of these structures depend on different 
characteristics of the materials and of the formation process. Another possibility to produce 
structured substrates is to change the functionality of the SAMs partially by irradiating a 
homogeneous SAM which is covered with a mask that carries the desired structure \cite{Zhao2002}. 
A third method is the so-called micro-contact printing \cite{Kumar1993} where samples, e.g., produced 
with lithography, are used as a mold for an elastomer stamp. The structure of the original is copied 
by stamping a ``thiol ink'' on gold covered substrates which chemisorbs there and forms a SAM. The 
empty spaces between the patterns of the stamped structure can be filled with a second thiol with a 
different functional end-group such that a topologically flat but chemically structured substrate is 
built up. As a last method we mention the exposure of oxidated titanium surfaces to UV-light 
\cite{Wang1997, Wang1998, Seki2004}. \\
\indent In order to investigate the structural properties of fluid systems near surfaces various experimental 
methods have been developed: Ellipsometry and especially phase modulated ellipsometry have been 
established in Refs. \cite{Jasperson1969} and \cite{Beaglehole1980} more than 20 years ago and are 
still powerful tools \cite{Cho2001, Howse2002}. An incident light beam is reflected by the surface of 
interest and the ratio of the complex reflection amplitudes for polarizations parallel and 
perpendicular to the plane of incidence (``coefficient of ellipsometry'' or ``ellipticity'') is 
measured. The order parameter is modeled and related to the ellipticity via the spatially varying 
dielectric constant and can then be compared with the measured value.
In neutron or X-ray reflectometry one measures the reflectivity as a function of the momentum 
transfer normal to the surface of the sample. This reflectivity spectrum is related to the refractive 
index profile which itself is associated with the profile of the order parameter \cite{Howse2002, 
Penfold1990, Thirtle1997, Dietrich1987, Geoghegan1998}.
A more direct measurement of the adsorption of a component to a substrate can be carried out with a 
differential refractometer \cite{Gruell1996, Rother2004}. Here a laser beam passes a measurement cell 
consisting of two compartments filled with the fluid of interest and a reference liquid, respectively. 
The intensity of the deflected beam is proportional to their difference in refractive index. By 
comparing this measured intensity with the one of a beam deflected by a measurement cell filled with 
a liquid with known interfacial properties and the reference liquid one is able to infer the properties 
of the fluid of interest.

With these methods critical adsorption on homogeneous substrates has been experimentally investigated 
(see Refs. \cite{Floeter1995} and \cite{Law2001} and references therein). On chemically structured 
substrates mostly wetting experiments have been performed \cite{Drelich1993, Gau1999, Geoghegan2003}. 
The present paper extends theoretical work on critical adsorption on chemically homogeneous 
\cite{Binder1983, Diehl1986, Fisher1978, Smith1997} and topologically \cite{Palagyi2004} structured 
surfaces and work on wetting phenomena at chemically structured substrates \cite{Bauer1999, Bauer1999a, 
Brinkmann2002, Schneemilch2003, Ez2004} to the case of critical adsorption on geometrically flat, but 
chemically structured surfaces.

The remainder of this paper is organized as follows: In Sec.\,\ref{sec:hom-sub} we introduce our model
and in order to set the stage we recall previous results on critical adsorption at homogeneous substrates. 
In Sec.\,\ref{sec:inhom-sub} we present our results for the critical adsorption at chemically structured 
substrates. We summarize our findings in Sec.\,\ref{sec:summary}.

\section{\label{sec:hom-sub}Critical adsorption at homogeneous substrates}
%
Boundaries, which come into play when investigating confined systems, induce deviations from the bulk 
behavior. Near a critical point the Ising bulk universality class splits into three possible surface 
universality classes (denoted as normal, special, and ordinary surface universality class, respectively) 
characterized by surface critical exponents and amplitudes \cite{Binder1983, Diehl1986}. The boundary 
conditions applied to the system determine the surface universality class the system belongs to. The 
behavior of the bulk is not affected by the boundaries.

It has turned out that in the sense of renormalization group theory it is sufficient to describe the 
presence of the substrate by a surface field $h_1$ and the so-called surface enhancement $c$ 
\cite{Diehl1986}. (These names originate from the corresponding description of surface magnetic 
phenomena \cite{Binder1983, Diehl1986}, see also, c.f., Eq.\,(\ref{eq:Ham-surf}).) The surface 
enhancement is related to the couplings between the ordering degrees of freedom at the surface. 

The so-called ordinary surface universality class is characterized by a vanishing surface field and 
a positive surface enhancement ($h_1=0, c>0$) which suppresses the order para\-meter at the surface below
its bulk value. For magnetic systems this effect of missing bonds is the generic case.

The special surface universality class, describing a multicritical point, requires in addition to a 
vanishing surface field a surface enhancement which within mean field theory vanishes ($h_1=0, c=0$)
and causes a flat order parameter profile in the vicinity of the substrate; fluctuations induce a 
divergence of the order parameter profile at the surface.

The normal surface universality class is characterized by a non-vanishing surface field and the 
absence of the surface enhancement ($|h_1|>0, c=0$) which leads to an order parameter value at the 
surface larger than in the bulk, even above the critical temperature where the bulk value of the 
order parameter is 0. For fluid systems the normal surface universality class is the generic 
case. In contrast, in the context of magnetism an order parameter which is larger at the surface than
its value in the bulk can be obtained in the absence of surface fields but for a negative surface 
enhancement ($h_1=0, c<0$). Since this is rather uncommon for magnetic systems, the normal surface 
universality class is also referred to as the so-called extraordinary surface universality class. 
The fact that the normal and the extraordinary cases are equivalent and identical at their fixed 
points, ($|h_1| \to \infty, c=0$) for fluid systems and ($h_1=0, c\to -\infty$) for 
magnetic systems, and thus identical with respect to their asymptotic behavior, was predicted by Bray 
and Moore \cite{Bray1977} and later proven by Burkhardt and Diehl \cite{Burkhardt1994}. In the 
following we shall focus on the normal and extraordinary surface universality classes, respectively.

A homogeneous substrate confining a binary liquid mixture inevitably has a preference for one of the 
two components. This preference becomes pronounced at the critical point and leads to an enrichment 
of the preferred component at the substrate. At the critical temperature the local order parameter 
profile decays algebraically towards its bulk value. This phenomenon has been called critical 
adsorption \cite{Fisher1978}.

In the case of planar homogeneous substrates besides the correlation length $\xi^{\pm}$ the distance
$z$ from the substrate is the other relevant length scale. If this length is scaled with the bulk
correlation length $\xi^{\pm}$, the order parameter profile $\phi(z,t)$ takes the following scaling 
form at the fixed points ($|h_1| \to \infty, c=0$) and ($h_1=0, c\to -\infty$), 
respectively:
  \begin{eqnarray}
    \phi(z,t) = a |t|^{\beta} P^{\pm}\bigg(w=\frac{z}{\xi^\pm}\bigg) \qquad \text{for } t \gtrless 0 \,,
    \label{eq:scalingbehavior-1d}
  \end{eqnarray}
with the scaling variable $w=z/\xi^\pm$ which describes the distance from the substrate in
units of the correlation length $\xi^\pm$.

The scaling functions $P^{\pm}(w)$ are universal after fixing the non-universal amplitude $a$ and 
the non-universal amplitude $\xi_0^+$ of the correlation length $\xi$. (Recall that the ratio 
$\xi_0^+ / \xi_0^-$ is universal and therefore the amplitude $\xi_0^-$ is fixed along with the 
amplitude $\xi_0^+$.) The amplitude $a$ is chosen in such a way that the scaling function $P^-(w)$ 
below the critical temperature tends to 1 for large distances from the substrate, i.e., the 
amplitude $a$ corresponds to the amplitude of the bulk order parameter 
$\phi(z \to \infty, t<0) = a |t|^{\beta}$. With this choice one finds the following behavior 
of the scaling functions:
  \begin{eqnarray}
    P^{\pm}(w \to 0) \sim w^{-\frac{\beta}{\nu}} \,,
    \label{eq:asymp_P-1} \\
    P^{+}(w \to \infty) \sim e^{-w} \,,
    \label{eq:asymp_P-2} \\
    P^{-}(w \to \infty) - 1 \sim e^{-w} \,.
    \label{eq:asymp_P-3} 
  \end{eqnarray}

Away from the renormalization group fixed point the surface field $h_1$ and the surface enhancement 
$c$ have finite values. They appear as additional parameters in the scaling functions with powers of 
the reduced temperature as prefactor due to scaling with the correlation length:
  \begin{eqnarray}
    \phi(z,t) &=& a |t|^{\beta} P^{\pm}(|t|^{\nu} \, ({\xi^\pm_0})^{-1} \, z,
                                        |t|^{-\Delta_1} \, (\xi^\pm_0)^{\frac{d}{2}} \, h_1, 
                                        |t|^{-\Phi} \, \xi^\pm_0 \, c) \,,
  \end{eqnarray}
where $\Delta_1$ and $\Phi$ are surface critical exponents \cite{Diehl1986} and $d$ denotes the spatial 
dimension of the system.

The explicit calculation of the order parameter profiles $\phi(z,t)$ starts from the following fixed
point Hamiltonian $\mathcal{H}[\phi]= \mathcal{H}_b[\phi]+ \mathcal{H}_s[\phi]$ which separates into 
the bulk part $\mathcal{H}_b[\phi]$ in the volume $V$ and the surface part $\mathcal{H}_s[\phi]$ on 
the surface $S$ \cite{Binder1983, Diehl1986}:
  \begin{eqnarray}
    \mathcal{H}_b [\phi] &=& \int_V \! \dd^{d-1} \mathbf{r_{\parallel}} \,\dd z 
                             \left( \2(\nabla \phi)^2 + \2 \tau \phi^2 + \frac{u}{4!} \phi^4 \right) \,,
    \label{eq:Ham-bulk} \\
    \mathcal{H}_s [\phi] &=& \int_S \! \dd^{d-1} \mathbf{r_{\parallel}} \left( \2 c \phi^2 - h_1 \phi \right) \,.
    \label{eq:Ham-surf}
  \end{eqnarray}
Here $\tau$ is proportional to the reduced temperature $t$, $u>0$ stabilizes the Hamiltonian 
$\mathcal{H}[\phi]$ for temperatures below the critical point ($T<T_c$) and $(\nabla \phi)^2$ 
penalizes spatial variations; $\mathbf{r_{\parallel}}$ is a vector parallel to the substrate.
The order parameter $\phi$ is fluctuating around a mean value $\langle \phi \rangle$. Each 
configuration $\phi$ contributes to the partition function $Z$ with the statistical Boltzmann weight 
$e^{-\mathcal{H}[\phi]}$:
  \begin{eqnarray}
    Z = \int \! \mathrm{D}\phi \left( e^{-(\mathcal{H}_b[\phi] + \mathcal{H}_s[\phi])} \right) \,, \\
    \langle \phi \rangle = \frac{1}{Z} \int \! \mathrm{D}\phi 
                           \left( \phi \, e^{-(\mathcal{H}_b[\phi] + \mathcal{H}_s[\phi])} \right) \,.
  \end{eqnarray}

In the present work we shall provide general scaling properties with the quantitative results for the
scaling functions determined within mean field approximation, i.e., only the order parameter profile 
$m(z)$ with the maximum statistical weight will be considered and all others will be neglected:
  \begin{eqnarray}
    \left. \frac{\delta \mathcal{H}[\phi]}{\delta \phi} \right|_{\phi = m} = 0 \,.
    \label{eq:mfa}
  \end{eqnarray}
This mean field approximation is valid above the upper critical dimension $d_c= 4$.  this 
approximation the aforementioned critical exponents take the following values:
  \begin{eqnarray}
    \beta (d \geq 4) = \2 \, \quad \text{and} \quad \nu(d \geq 4) = \2 \,,
    \label{eq:expo-d=4}
  \end{eqnarray}
whereas the critical exponents at physical dimension $d=3$ are \cite{LeGuillou1985}:
  \begin{eqnarray}
    \beta (d=3) = 0.3265 \, \quad \text{and} \quad \nu(d=3) = 0.6305 \,.
    \label{eq:expo-d=3}
  \end{eqnarray}
The mean field approximation is important because it is the zeroth-order approximation in
a systematic Feynman graph expansion on which the $(\epsilon=d-4)$-expansion and hence the 
renormalization group approach are based \cite{Zinn-Justin1989, Diehl1986, Brezin1976}. The higher 
orders in the Feynman graph expansion require integrations over the mean field order parameter profile 
and the two-point correlation function (compare Eq.\,(3.209) in Ref. \cite{Diehl1986}) which can be 
carried out reasonably if they are available in an analytical form. However, the mean field 
approximation is expected to yield the qualitatively correct behavior of the scaling functions if for
the variables forming the scaling variables the correct critical exponents are used, which are known 
with high accuracy \cite{LeGuillou1985} (see Eq.\,(\ref{eq:expo-d=3})).

Taking the functional derivative of the Hamiltonian with respect to the order parameter (see Eqs. 
(\ref{eq:Ham-bulk}), (\ref{eq:Ham-surf}), and (\ref{eq:mfa})) 
yields a differential equation for the mean field profile $m(z)$ of the order parameter 
\cite{Binder1983, Diehl1986},
  \begin{eqnarray}
    - \frac{\partial^2}{\partial z^2} m + \tau m + \frac{u}{3!} m^3 = 0 \,,
    \label{eq:diff-eq} 
  \end{eqnarray}
with boundary conditions
  \begin{eqnarray}
    \left. \frac{\partial m}{\partial z} \right|_{z=0} = c \, m(z=0) - h_1
    \label{eq:bound-cond-sub}
  \end{eqnarray}
and
  \begin{eqnarray}
    \left. \frac{\partial m}{\partial z} \right|_{z \to \infty} = 0 \,.
    \label{eq:bound-cond-bulk}
  \end{eqnarray}

\setcounter{subsubsection}{0}
\subsubsection{Infinite surface fields}
%
For systems that belong to the extraordinary surface universality class ($h_1 = 0, c<0$) the 
boundary condition at the surface (\ref{eq:bound-cond-sub}) simplifies and the differential equation 
(\ref{eq:diff-eq}) has an analytical solution \cite{Diehl1986}. Together with the scaling behavior 
of the order parameter (\ref{eq:scalingbehavior-1d}) and the non-universal amplitude $a$ which within 
the present model and within mean field (MF) approximation equals 
$a=\left( \frac{6}{u} \right)^{\frac{1}{2}} \, \left( \xi^\pm_0 \right)^{-1}$ this leads 
to scaling functions $P^{\pm}_{MF}(w)$ of the following form:
  \begin{eqnarray}
    P^+_{MF}(w) &=& \frac{\sqrt{2}}{\sinh (w+w_0)} \, ; \quad \coth(w_0) = |\tilde{c}| \,,
    \label{eq:solplus-diffeq} \\
    P^-_{MF}(w) &=& \coth \left(\frac{w+w_0}{2}\right) \, ; \quad \sinh(w_0) = \frac{1}{|\tilde{c}|} \,,
    \label{eq:solminus-diffeq}
  \end{eqnarray}
where $\tilde{c}=\xi^{\pm}_0 |t|^{-\frac{1}{2}} c$ is the scaled and dimensionless surface enhancement.
The parameter $w_0$ vanishes at the extraordinary fixed point ($\tilde{h}_1=0, \tilde{c}\to -\infty$) 
-- with the scaled and dimensionless surface field 
$\tilde{h}_1 = \left( \frac{u}{6} \right) ^{\frac{1}{2}} \left.\xi^\pm_0\right.^2 |t|^{-1} h_1$ -- so that Eqs. 
(\ref{eq:solplus-diffeq}) and (\ref{eq:solminus-diffeq}) reduce to the scaling functions
  \begin{eqnarray}
     P^+_\infty(w) = \frac{\sqrt{2}}{\sinh(w)}
    \label{eq:halfspaceprof_plus}
  \end{eqnarray}
and
  \begin{eqnarray}
     P^-_\infty(w) = \coth \left({\textstyle \frac{w}{2}} \right) \,,
    \label{eq:halfspaceprof_minus}
  \end{eqnarray}
which in the following are referred to as ``half-space profiles'' above and below the critical temperature,
respectively. As already mentioned at the beginning of Sec.\,\ref{sec:hom-sub} the extraordinary fixed 
point is equivalent to the normal fixed point ($|\tilde{h}_1| \to \infty, \tilde{c}=0$) and
thus Eqs.\,(\ref{eq:halfspaceprof_plus}) and (\ref{eq:halfspaceprof_minus}) represent the latter 
fixed point as well.

\subsubsection{\label{sec:fin_field}Finite surface fields}
%
Since in experimental systems the surface fields are finite, we also discuss the case of a 
homogeneous substrate with a surface field $0<h_1< \infty$. This provides the starting point for 
discussing the case of a chemical step with finite - albeit strong - surface fields which we shall 
consider in Subsec. \ref{subsubsec:chemstep-asym}. Off criticality (i.e., $t\neq0$) and for a finite 
surface field $h_1$ the scaled surface field 
$\tilde{h}_1 = \left( \frac{u}{6} \right) ^{\frac{1}{2}} \left.\xi^\pm_0\right.^2 |t|^{-1} h_1$ is also 
finite.
At the substrate, for finite surface fields the order parameter profiles $P^\pm_{h_1}(w)$ above/below 
the critical temperature have finite values $P^+_{h_1}(w=0)=P^+_{sub}$ and $P^-_{h1}(w=0)=P^-_{sub}$, 
respectively. These values are determined by the following equations which are obtained by performing 
the first integral of the differential equations for $P^+_{MF}$ and $P^-_{MF}$, corresponding to 
Eq.\,(\ref{eq:diff-eq}) \cite{Binder1983}:
  \begin{eqnarray}
    \left.P^+_{sub}\right.^4 + 2(1-\tilde{c}^2)\, \left.P^+_{sub}\right.^2 
    + 4 \tilde{c}\, \tilde{h}_1\, P^+_{sub} - 2 \tilde{h}_1^2 = 0
    \label{eq:master_P+}
  \end{eqnarray}
and
  \begin{eqnarray}
    \left.P^-_{sub}\right.^4 + 2(1+2\tilde{c}^2)\, \left.P^-_{sub}\right.^2 
    - 8 \tilde{c}\, \tilde{h}_1\, P^-_{sub} + 4 \tilde{h}_1^2 = 0 \,.
    \label{eq:master_P-}
  \end{eqnarray}
   \begin{figure}
    \begin{center}
      \epsfig{file=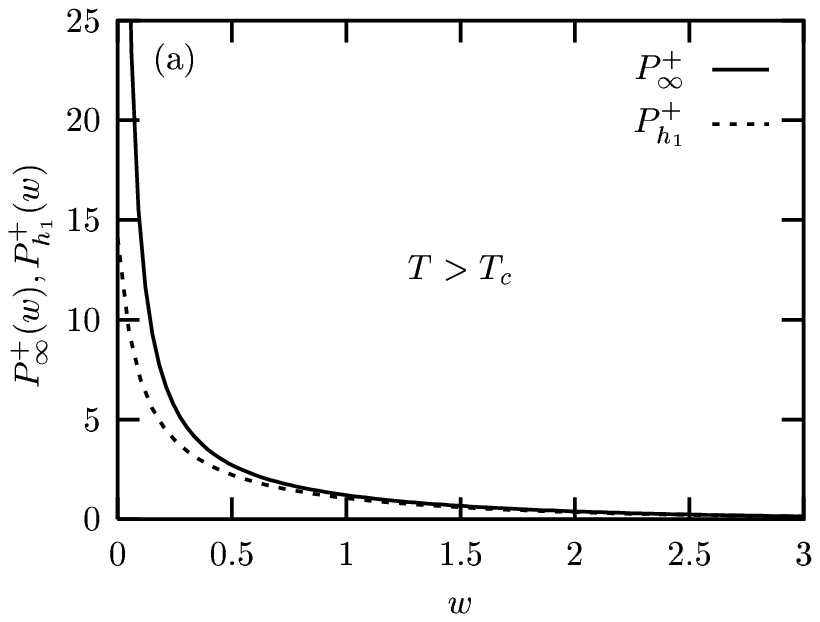,width=0.5\textwidth} \quad
      \epsfig{file=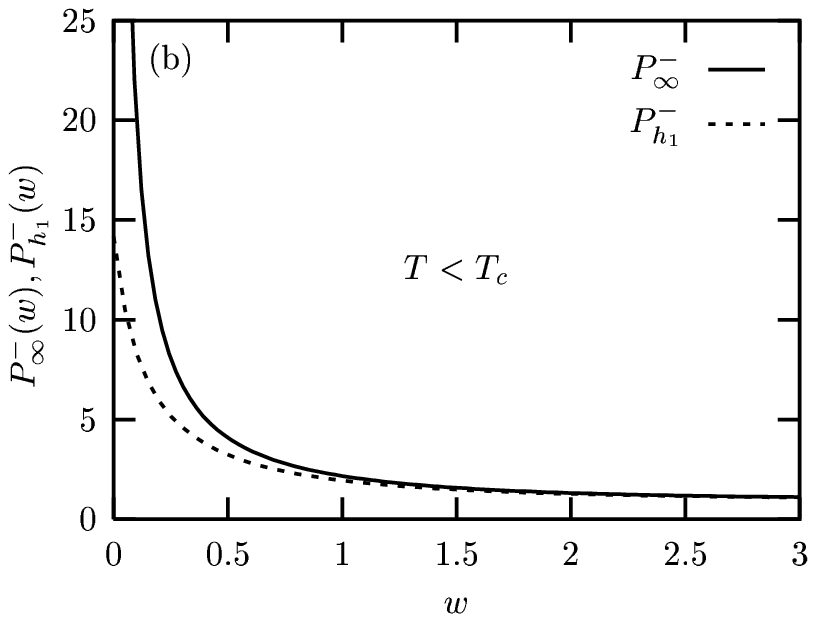,width=0.5\textwidth}
      \caption{\label{fig:half-space}(a) The half-space profile $P^+_\infty(w=z/\xi^+)$ for an infinite surface 
               field $\tilde{h}_1$ and $P^+_{h_1}(w=z/\xi^+)$ for a finite scaled surface field $\tilde{h}_1=100$ 
               above the critical temperature. (b) The same below $T_c$ with $w=z/\xi^-$.}
    \end{center}
  \end{figure}
\hspace*{-2mm}For the normal surface universality class (i.e., $\tilde{c}=0$) Eqs.\,(\ref{eq:master_P+}) and 
(\ref{eq:master_P-}) simplify and we find for the order parameter $P^+_{sub}$ and $P^-_{sub}$ at the 
substrate above and below the critical temperature, respectively:
  \begin{eqnarray}
    P^+_{sub}= \pm \sqrt{\sqrt{1 + 2 \tilde{h}_1^2}-1} \ , \quad \tilde{h}_1 \lessgtr 0
  \end{eqnarray}
and
  \begin{eqnarray}
    P^-_{sub}= \pm \sqrt{\sqrt{1 + 4 \tilde{h}_1^2}+1} \ , \quad \tilde{h}_1 \lessgtr 0 \,.
    \label{eq:psub}
  \end{eqnarray}
Using these equations together with the boundary condition (\ref{eq:bound-cond-sub}) and the 
differential equation (\ref{eq:diff-eq}) the half-space profile for finite surface fields 
$P^{\pm}_{h_1}(w) = P^{\pm}(w, \tilde{h}_1 < \infty, \tilde{c}=0)$ can be calculated numerically
(see Appendix \ref{app}). In Fig.\,\ref{fig:half-space} the half-space profiles for 
infinite and finite surface fields are shown for temperatures above as well as below the critical
temperature. In the following section we shall consider inhomogeneous substrates, i.e., substrates 
with a laterally varying surface field $h_1$ for which further length scales come into play, which 
also scale with the correlation length $\xi^\pm$.%

\section{\label{sec:inhom-sub}Critical adsorption at inhomogeneous substrates}
%
\subsection{\label{subsec:chem_step}Chemical step}
%
First we consider an infinite substrate which is divided into two halves with opposing surface 
fields $h_1$ so that there is a chemical step at the straight contact line ($x=z=0$) of the two 
halves (see Fig.\,\ref{fig:geometry}(a)).
We introduce the scaled coordinates $v=x/\xi$ and $w=z/\xi$ describing the distance $x$ from the 
contact line and $z$ from the substrate, respectively, in units of the correlation length $\xi$. 
The system is translationally invariant in the direction perpendicular to the $x$-$z$ -- plane.

For laterally inhomogeneous systems one has to reconsider whether the surface Hamiltonian 
$\mathcal{H}_s$ (Eq.\,(\ref{eq:Ham-surf})) should contain terms like $\phi \partial_{\parallel} \phi$ and 
$\phi \partial_{\perp} \phi$. However, the term $\phi \partial_{\parallel} \phi$ would favor order 
parameter profiles which are non-symmetric with respect to $(x=0,y)$ even without surface fields. Therefore
such a term is ruled out. The term $\phi \partial_{\perp} \phi$ leads only to a redefinition of the 
surface enhancement \cite{Diehl1986} and therefore can be neglected for homogeneous as well as for 
the inhomogeneous substrates. Thus the surface Hamiltonian $\mathcal{H}_s$ for inhomogeneous surface 
fields has the same form as the one for homogeneous surface fields (Eq.\,(\ref{eq:Ham-surf})).

\subsubsection{Infinite surface fields}
%
First we analyze the case of a homogeneous infinite surface field on both halves of the substrate 
but with opposite sign and a vanishing surface enhancement, i.e., we consider a step-like lateral
variation of the surface field $h_1$: $h_1 = \pm \infty$ for $x \gtrless 0$. The 
actual smooth variation of $h_1$  on a microscopic scale turns effectively into a step-like variation 
if considered on the scale $\xi^\pm$.

\paragraph{Order parameter profiles}
\hspace*{0mm}

\begin{figure}
    \begin{center}
      \epsfig{file=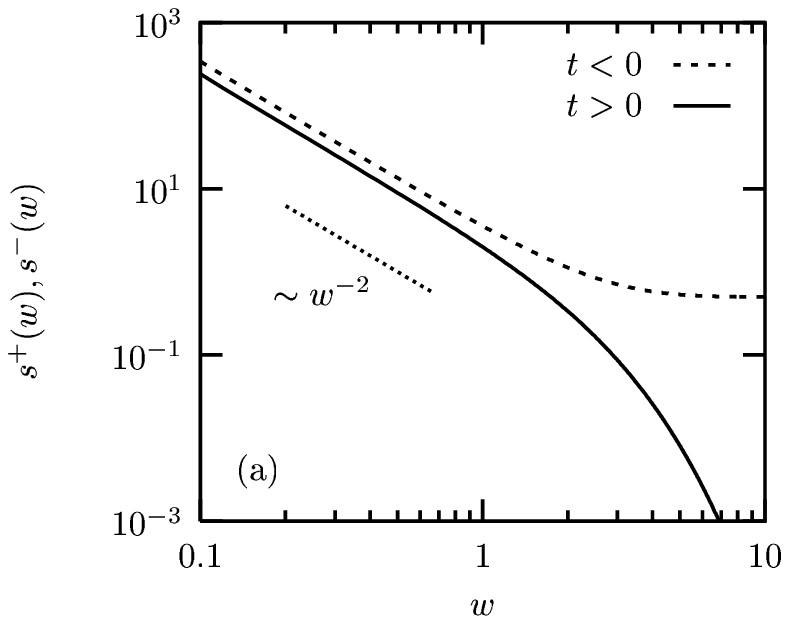,width=0.5\textwidth} \quad
      \epsfig{file=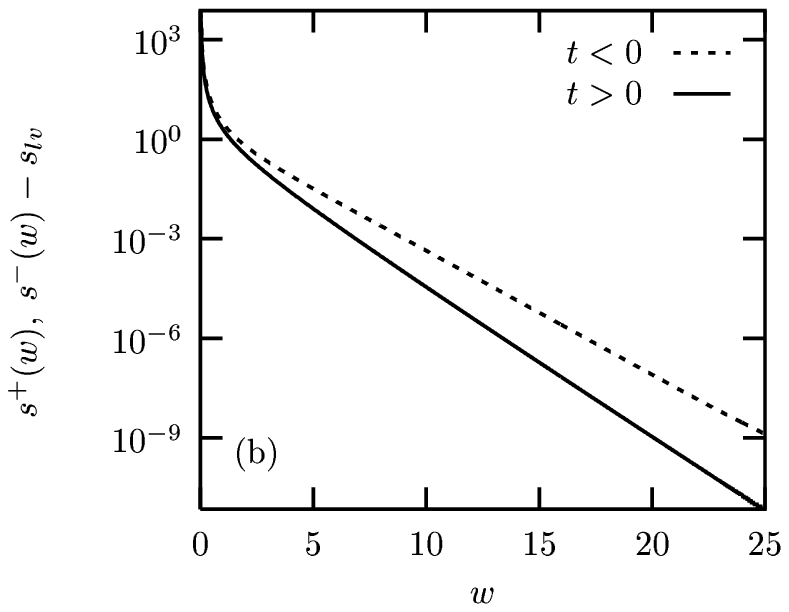,width=0.5\textwidth}
      \caption{\label{fig:slope}Slopes $s^\pm(w)$ at the chemical step of the scaling functions 
               $P^\pm_{cs}(v=x/\xi^\pm, w=z/\xi^\pm)$ for the order parameter profile of a system 
               with a chemical step at $x=0$. (a) For small distances from the substrate the 
               slopes diverge as $A^\pm_1 w^{-2}$ with amplitudes $A^+_1=2.367 \pm 0.006$ and
               $A^-_1=3.366 \pm 0.005$ (see Eq.\,(\ref{eq:s_pm-w=0})).
               (b) For large distances the slopes decay exponentially, above $T_c$ \,
               $s^+(w \to \infty) = A^+_2 e^{-w}$ with $A^+_2=0.566 \pm 0.001$ (see Eq. 
               (\ref{eq:s_p-w=infty})); below $T_c$ \, $s^-(w \to \infty)$ approaches its limiting
               value $s_{lv}=\2$ slower, i.e., as $A^-_2 e^{-Cw}$ with $A^-_2=2.338 \pm 0.001$ and 
               $C=0.858 \pm 0.001$ (see Eq.\,(\ref{eq:s_m-w=infty})).}
    \end{center}
  \end{figure}

The order parameter profile for a system with a chemical step (cs) exhibits the following scaling property
  \begin{eqnarray} 
    \phi(x,z,t) = a |t|^\beta P^\pm_{cs}\bigg(v=\frac{x}{\xi^\pm}, w=\frac{z}{\xi^\pm}\bigg) \quad \text{for } t \gtrless 0\,
    \label{eq:scalingbehavior-2d}
  \end{eqnarray}
which generalizes Eq.(\ref{eq:scalingbehavior-1d}). This scaling function shows the following limiting 
behavior: For large distances from the chemical step ($|v| \to \infty$) the profile approaches 
the corresponding order parameter profile of a system with a homogeneous substrate, whose asymptotic 
behavior is given by Eqs.\,(\ref{eq:asymp_P-1})-(\ref{eq:asymp_P-3}), 
$P^\pm_{cs}(|v| \to \infty, w) = P^\pm_\infty(w)$. For large distances from the substrate 
($w\to \infty$) and above the critical temperature the order parameter profile vanishes for all values 
of $v$, i.e., $P^+_{cs}(v,w \to \infty) = 0$ while below the critical temperature the order parameter 
profile tends to the profile $P^-_{lv}(v)$ of a free liquid-vapor interface 
$P^-_{cs}(v,w\to \infty) = P^-_{lv}(v)$ \cite{liquid-vapor}.  the mean field approximation this 
profile is given by
   \begin{eqnarray}
     P^-_{lv} (v)=  \tanh \left( \frac{v}{2} \right) \,.
     \label{eq:lv-interface}
   \end{eqnarray}
  \begin{figure}
    \begin{center}
      \epsfig{file=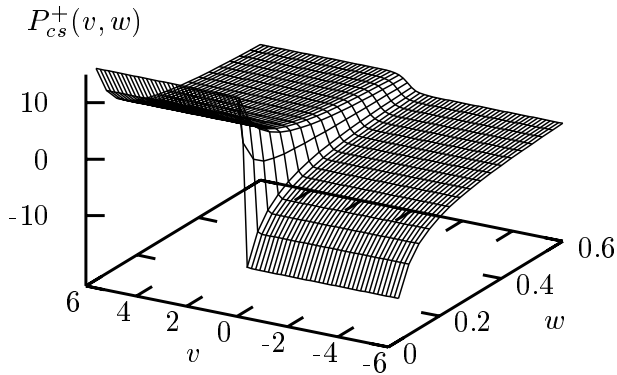,width=0.5\textwidth}
      \caption{\label{fig:3d-plot} Scaling function $P^+_{cs}(v=x/\xi^+, w=z/\xi^+)$ for the 
               order parameter profile of a system above the critical temperature and confined 
               by a substrate with a chemical step located at $x=z=0$. Due to symmetry 
               $P^+_{cs}(v=0,w) = 0$.}
    \end{center}
  \end{figure}
\indent Figure\,\ref{fig:3d-plot} provides a three-dimensional plot of the numerically determined 
(see Appendix \ref{app}) order parameter profile for $T>T_c$. In Fig.\,\ref{fig:cross} we 
show cuts through the order parameter profile parallel to the substrate as it changes with 
increasing distance from the substrate.
  \begin{figure}
    \begin{center}
      \epsfig{file=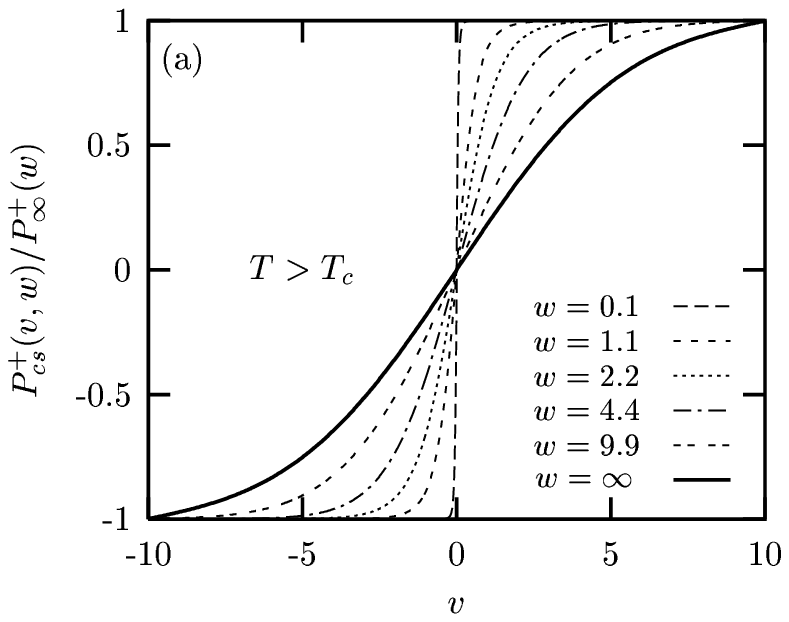,width=0.5\textwidth} \quad
      \epsfig{file=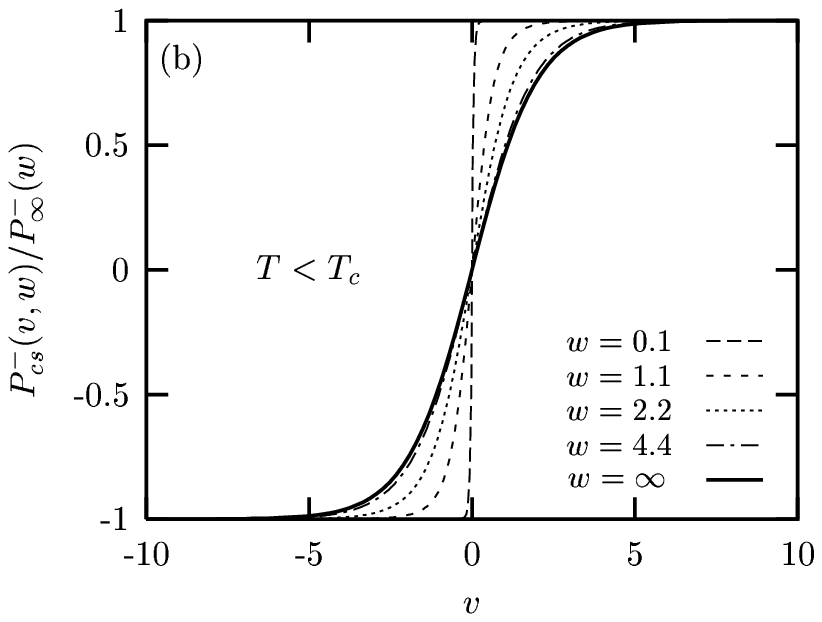,width=0.5\textwidth}
      \caption{\label{fig:cross}Cuts through the order parameter profiles $P^+_{cs}(v,w)$, as shown 
               in Fig.\,\ref{fig:3d-plot}, and $P^-_{cs}(v,w)$ for $w=z/\xi=\mathit{const}$ with the 
               normalization $\frac{P^\pm_{cs}(v,w)}{P^\pm_\infty(w)}$ (a) above and (b) below the 
               critical temperature. Upon increasing the distance from the substrate the slope of 
               the curves at $v=0$ decreases. The solid lines represent the limiting curves for 
               $w\to \infty$: (a) Above the critical temperature the limiting slope at $v=0$ of the 
               normalized curves is given by 
               $\frac{s^+(w \to \infty)}{ P^+_\infty(w \to \infty)} = \frac{A^+_2}{2\sqrt{2}} \simeq 0.200$
               with the slope $s^+(w) = \left. \frac{\partial P^+_{cs}(v,w)}{\partial v} \right|_{v=0}$
               of the unnormalized scaling function and its amplitude $A^+_2 \simeq 0.566$ (see, c.f.,
               Eq.\,(\ref{eq:s_p-w=infty}) and Fig.\,\ref{fig:slope}). (b) Below $T_c$ the limiting 
               curve corresponds to $\tanh(\frac{v}{2})$.}
    \end{center}
  \end{figure}
In order to provide a clearer comparison the cross sections are normalized to 1 at the lateral 
boundaries via dividing by the exponentially decaying half-space profile $P^\pm_\infty(w)$. From the
fact that the cross sections do not fall onto one curve follows that the scaling function 
$P^\pm_{cs}(v,w)$ does not separate into a $v$-dependent and a $w$-dependent part. It shows that 
the slope $s^\pm(w)= \left. \frac{\partial P^\pm_{cs}(v,w)}{\partial v} \right|_{v=0}$ of the
scaling function at the step decreases with increasing distance from the substrate which visualizes
the healing of the frustration upon approaching the bulk. Note that the slopes in Fig.\,\ref{fig:cross}
must be multiplied by $P^\pm_\infty(w)$ in order to obtain $s^\pm(w)$.
From Eq.\,(\ref{eq:scalingbehavior-2d}) it follows:
  \begin{eqnarray}
    \left. \frac{\partial \phi(x,z,t)}{\partial x} \right|_{x=0}
    &=& a \frac{|t|^{\beta}}{\xi} \left. \frac{\partial P^{\pm}_{cs}(v,w)}{\partial v} \right|_{v=0} 
        \nonumber \\
    &=& \frac{a}{\xi^\pm_0} \, |t|^{\beta +\nu} s^{\pm}(w) \ .
    \label{eq:powers}
  \end{eqnarray}
Since there is a non-vanishing order parameter profile even at the critical point ($\phi(x,z,t=0)\neq0$) 
the overall temperature dependence of the slope $\frac{\partial \phi}{\partial x}$ in Eq.\,(\ref{eq:powers}) 
has to vanish. Therefore one finds for $w= \frac{z}{\xi^\pm_0} t^\nu \to 0$:
   \begin{eqnarray}
      s^{\pm}(w \to 0) = A^\pm_1 w^{-\frac{\beta}{\nu}-1} \,.
      \label{eq:s_pm-w=0}
   \end{eqnarray}
For $T>T_c$ the slope $s^+(w)$ vanishes exponentially upon approaching the bulk:
   \begin{eqnarray}
     s^+(w \to \infty) = A^+_2 e^{-w} \ .
     \label{eq:s_p-w=infty}
   \end{eqnarray}
Below $T_c$ the slope $s^-(w \to \infty)$ approaches the slope 
$s_{lv} = \left. \frac{\partial P^-_{lv}(v)}{\partial v} \right|_{v=0}$ of the scaling function for 
the liquid-vapor profile (Eq.\,(\ref{eq:lv-interface})) at most $\sim e^{-w}$ or slower:
   \begin{eqnarray}
     s^-(w\to \infty) - s_{lv} = A^-_2 e^{-Cw} \,, \quad 0 < C \leq1 \,.
     \label{eq:s_m-w=infty}
   \end{eqnarray}
As the half-space profiles $P^+_{cs}(v \to \pm \infty, w) = \pm P^+_\infty(w)$ (Eq.\,(\ref{eq:asymp_P-2}))
and $P^-_{cs}(v \to \pm \infty, w) = \pm P^-_\infty(w)$ (Eq.\,(\ref{eq:asymp_P-3})) decay 
exponentially $\sim e^{-w}$ for large distances from the substrate, the slope $s^\pm(w)$ cannot decay 
faster because this would require that the slope $s^\pm(w)$ is smaller than the slope 
$\left. \frac{\partial P^+_{cs}}{\partial v} \right|_{v_0}$ for a certain $v_0 \neq 0$. On the other hand a 
decay of the slope $s^+(w)$ slower than $e^{-w}$ would lead to an unphysical increasing slope of the 
normalized cross sections at $v=0$ with increasing distance from the substrate. However, for the slope 
$s^-(w)$ a decay towards $s_{lv}$ slower than $\sim e^{-w}$ cannot be ruled out. \\
\indent Within mean field theory Eqs.\,(\ref{eq:s_pm-w=0}) and (\ref{eq:lv-interface}) yield 
$s^{\pm}(w \to 0) \sim w^{-2}$ (see also Eq.\,(\ref{eq:expo-d=4})) and $s^-(w \to \infty) = s_{lv} = \2$, 
respectively, which is in agreement with the numerical results shown in Fig.\,\ref{fig:slope}.
These results for the slope $s^\pm(w)$ of the scaled order parameter profile $P^\pm_{cs}(v,w)$ transform 
into the following findings for the slope of the unscaled order parameter profile $\phi(x,z)$:
   \begin{eqnarray}
     \left. \frac{\partial \phi}{\partial x} \right|_{x=0} 
     &\sim& z^{-\frac{\beta}{\nu}-1} \,, \quad T=T_c \,, \\
     \left. \frac{\partial \phi}{\partial x} \right|_{x=0}
     &\sim& t^{\beta+\nu} \, e^{-\frac{z}{\xi^+}} \,, \quad T > T_c \,, \quad z \gg \xi^+ \,, \nonumber \\
     &\sim& \left(\xi^+\right)^{-\frac{\beta}{\nu}-1}\,e^{-\frac{z}{\xi^+}} \,, 
   \end{eqnarray}
and
   \begin{eqnarray}
     \left. \frac{\partial \phi}{\partial x} \right|_{x=0} - \frac{\partial \phi_{lv}}{\partial x}
      &\sim& |t|^{\beta+\nu} \, e^{-C\frac{z}{\xi^-}} \,, \quad T < T_c \,, \quad z \gg \xi^+ \,, \nonumber \\
      &\sim& |\xi|^{-\frac{\beta}{\nu}-1}\,e^{-C\frac{z}{\xi^-}} \,.
   \end{eqnarray}

\begin{figure}
    \begin{center}
      \epsfig{file=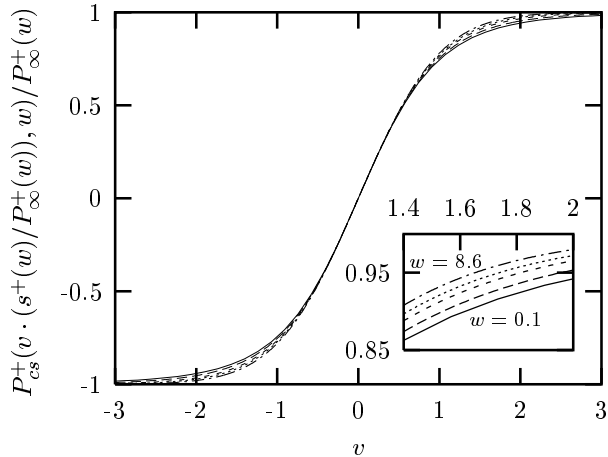,width=0.5\textwidth}
      \caption{\label{fig:slope-norm}Rescaled normalized cross sections with slope 1 at $v=0$ for
               $w=0.1, 1.5, 2.9, 4.3, 8.6$.}
    \end{center}
  \end{figure}

As an example we investigate more closely the healing effect above the critical temperature. To this 
end we rescale the normalized cross sections of Fig.\,\ref{fig:cross} such that their slope at $v=0$ 
becomes 1, see Fig.\,\ref{fig:slope-norm}.
It shows that these rescaled normalized cross sections for different distances from the substrate 
differ basically only in the region where the curvature of these curves is largest. From the fact 
that these cross sections do not fall onto one curve it follows that the scaling function $P^\pm(v,w)$
is not simply given by the knowledge of one cross section, the slope $s(w)$, and the half space profile
$P^+_{\infty}(w)$ but requires the full numerical analysis. Nonetheless Fig.\,\ref{fig:slope-norm} 
demonstrates that to a large extent the rescaling used there reduces the full scaling function 
$P^+_{cs}(v,w)$ to a single function of $v$ only.\\

\paragraph{Excess adsorption}
\hspace*{1mm}

It is a challenge to determine experimentally the full order parameter profile $\phi(x,z)$. Therefore in
the following we analyze the adsorption $\Gamma = \int \dd V \phi({\mathbf{r}})$ at the substrate 
which as an integral quantity is more easily accessible to experiments. To this end for any system 
with an order parameter profile $\phi(x,z)$ and a corresponding scaling function $P^\pm(v,w)$
we introduce a suitable reference system with an order parameter profile $\phi_{\mathit ref}(x,z)$ and the
corresponding scaling function $P_{\mathit ref}(v,w)$. This allows us to introduce an excess adsorption 
$\Gamma_{ex}$ with respect to this reference system
  \begin{eqnarray}
     \Gamma_{ex} &=& H \iint \dd x \,\dd z \left( \phi(x,z) - \phi_{\mathit ref}(x,z) \right)  \nonumber \\
                 &=& a |t|^\beta \left.\xi^\pm\right.^2 H \tilde{\Gamma}^{\pm}_{ex}  
     \label{eq:gamma_ex} 
  \end{eqnarray}
with its universal part $\tilde{\Gamma}_{ex}^\pm$ defined as (see Eqs.\,(\ref{eq:scalingbehavior-1d}) and 
(\ref{eq:scalingbehavior-2d}))
  \begin{eqnarray}
     \tilde{\Gamma}^{\pm}_{ex} &=& \iint \dd v \,\dd w \left( P^\pm(v,w) - P^\pm_{\mathit ref}(v,w) \right) \,.
     \label{gamma-tilde_ex}  
  \end{eqnarray} 
$H$ denotes the extension of the system perpendicular to the $x$-$z$ -- plane. (In the three-dimensional 
case $H$ corresponds to the one-dimensional extension of the system in $y$-direction. Within mean field 
theory, which is valid for dimensions $d\geq4$, $H$ corresponds to the $d-2$-dimensional extension of 
the system in the $y_1$-$\ldots$-$y_{d-2}$-directions.)
Equation (\ref{eq:gamma_ex}) leads to the following temperature dependence of the excess adsorption:
   \begin{eqnarray}
     \Gamma_{ex} = \tilde{\Gamma}^{\pm}_{ex} \, a \left.\xi^\pm_0\right.^2 \, H \,|t|^{\beta - 2 \nu} \,,
     \label{eq:gamma_ex-2}
   \end{eqnarray}
with the universal amplitude $\tilde{\Gamma}^{\pm}_{ex}$ and three non-universal amplitudes $a$, 
$\xi^\pm_0$, and $H$. Within mean field approximation this yields 
$\Gamma_{ex} = \tilde{\Gamma}^{\pm}_{ex} \, a \left. \xi^\pm_0\right.^2 \, H \, |t|^{-\frac{1}{2}}$.
For our choices of reference systems as given below $P^\pm_{\mathit ref}(v,w)$ leads even within mean 
field theory to a cancellation of the divergence of the corresponding integrals over the scaling 
function $P^\pm(v,w)$ caused by small distances from the substrate. This way the numerical mean field 
data for $d=4$ allow one to make meaningful approximate contact with potential experimental data for 
$d=3$.

\begin{figure}
    \begin{center}
      \epsfig{file=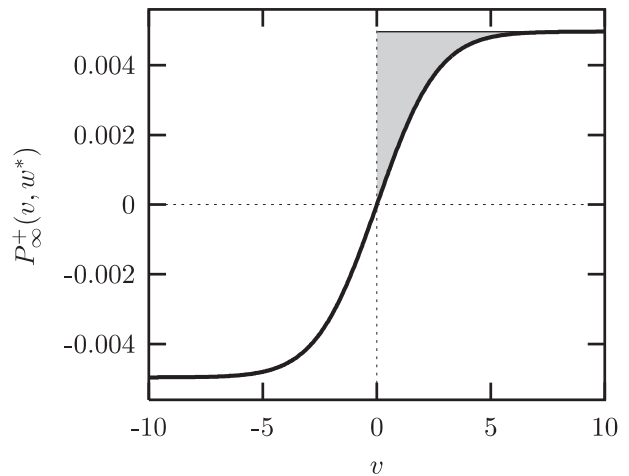,width=0.45\textwidth}
      \caption{\label{fig:integration}The shaded area indicates the contribution to the excess adsorption
               $\tilde{\Gamma}^{\pm}_{ex,cs}$ as defined in  Eq.\,(\ref{eq:integ-ex}) for $w=w^*=6.0$ 
               for a system above $T_c$ and with a chemical step.
               $\tilde{\Gamma}^{\pm}_{ex,cs}$ is obtained by summing these shaded areas over $w$.}
    \end{center}
  \end{figure}

Specifically, for the chemical step we introduce a reference system $P^{\pm}_{cs,0}(v,w)$ 
which can be interpreted as a system with a chemical step for which no healing at the chemical step 
occurs:
  \begin{eqnarray} 
    P^\pm_{cs,0}(v,w) = \left \{
              \begin{array}{rl}
                - P^\pm_\infty(w)\quad, & v<0 \\[2mm]
                + P^\pm_\infty(w) \quad, & v>0 \,.
              \end{array}
              \right.
    \label{eq:P_cs0}
  \end{eqnarray}
Since this choice of the reference systems leads to a vanishing excess adsorption for any antisymmetric 
(with respect to $v=0$) scaling function $P^\pm_{cs}(v,w)$ upon integration over the whole half space 
$w>0$, we restrict the integration to a quarter of the space (see Fig.\,\ref{fig:integration}):
   \begin{eqnarray}
     \tilde{\Gamma}^{\pm}_{ex,cs} = \int_0^\infty \! \dd v \int_0^\infty \! \dd w 
                                    \left( P^\pm_{cs}(v,w) - P^{\pm}_{cs,0}(v,w) \right) \,.
     \label{eq:integ-ex}
   \end{eqnarray}
\hspace*{-2mm}For a binary liquid mixture confined by a substrate with a chemical step $\Gamma^\pm_{ex,cs}$ can be 
interpreted as the amount of particles of one type removed across the chemical step from the 
substrate which prefers them.

Below $T_c$ the scaling function $P^-_{lv}(v)=P^-_{cs}(v,w=\infty)$ (Eq.\,(\ref{eq:lv-interface})) of 
the liquid-vapor profile gives rise to a non-vanishing universal excess adsorption 
$\tilde{\Gamma}_b$ with respect to the chosen reference system $P^-_{cs,0}$:
 \begin{eqnarray} 
    \tilde{\Gamma}_b 
    &=& \int_0^\infty \! \dd v \left( P^-_{lv}(v) - P^-_{cs,0}(v,w=\infty) \right) \\
    &=& \int_0^\infty \! \dd v \left (P^-_{lv}(v) - 1 \right) \nonumber \\
    &=& - 2 \ln 2 \,. \nonumber
  \end{eqnarray}
Thus $\tilde{\Gamma}^-_{ex,cs}$ can be written as 
  \begin{eqnarray} 
    \tilde{\Gamma}^-_{ex,cs} = \tilde{L} \, \tilde{\Gamma}_b + \tilde{\Gamma}^-_{ex,f} \,
    \label{eq:gammacs_gammabulk}
  \end{eqnarray}
where $\tilde{L}= L / \xi^-_0 \to \infty$ denotes the extension of the system perpendicular to the substrate 
and the system size independent contribution $\tilde{\Gamma}^-_{ex,f}$ characterizes the influence of 
the chemical step. From our numerical analysis we find the following universal excess adsorption 
amplitudes:
  \begin{eqnarray} 
   \tilde{\Gamma}^+_{ex,cs} &=& -1.457 \pm 0.001 \,,
   \label{eq:gammacs_plus} \\
   \tilde{\Gamma}^-_{ex,f} &=& 1.299 \pm 0.001 \,.
   \label{eq:gammacs_minus}
  \end{eqnarray}

\subsubsection{\label{subsubsec:chemstep-asym}Non-antisymmetric finite surface fields}
%
Next we study an infinite substrate consisting of two halves with surface fields of opposite sign but
different absolute finite values $h_p$ and $h_n$, respectively, and a vanishing surface enhancement. 
As before we consider a step-like variation of the surface field $h_1$:
  \begin{eqnarray}
     h_1(x) = \left \{
              \begin{array}{rl}
                h_n \quad, & x<0 \\[1mm]
                h_p \quad, & x>0 \,.
              \end{array}
              \right.
     \label{eq:hp-hn} 
  \end{eqnarray}
The absence of antisymmetry for these systems is reflected by the ``zero-line'' $v_0(w)$, where the 
order parameter vanishes: $P(v_0,w)=0$. For ratios $-\tilde{h}_p / \tilde{h}_n \neq 1$ of the scaled 
surface fields (see Subsec. \ref{sec:fin_field}) the zero-line is shifted towards the region of the 
surface field with the smaller absolute value. Furthermore the zero-line is not straight, but tends 
to increasing values $|v_0|$ for increasing distance $w$ from the substrate. However, the deviation 
of the zero-line from the line $(v=0,w)$ decreases for $-\tilde{h}_p / \tilde{h}_n \to 1$. For constant 
ratios $-\tilde{h}_p / \tilde{h}_n$ the deviation of the zero-line from the line $(v=0,w)$ decreases 
with increasing absolute values of the scaled surface fields $\tilde{h}_p$ and $\tilde{h}_n$.
These results demonstrate how the order parameter structure for the fixed point fields $|\tilde{h}_p|$, 
$|\tilde{h}_n| \to \infty$ emerges smoothly from the general case of finite fields. In this 
sense in the following we focus on the case of infinite surface fields, i.e., strong adsorption.

\subsection{Single stripe}
\begin{figure}
    \begin{center}
      \epsfig{file=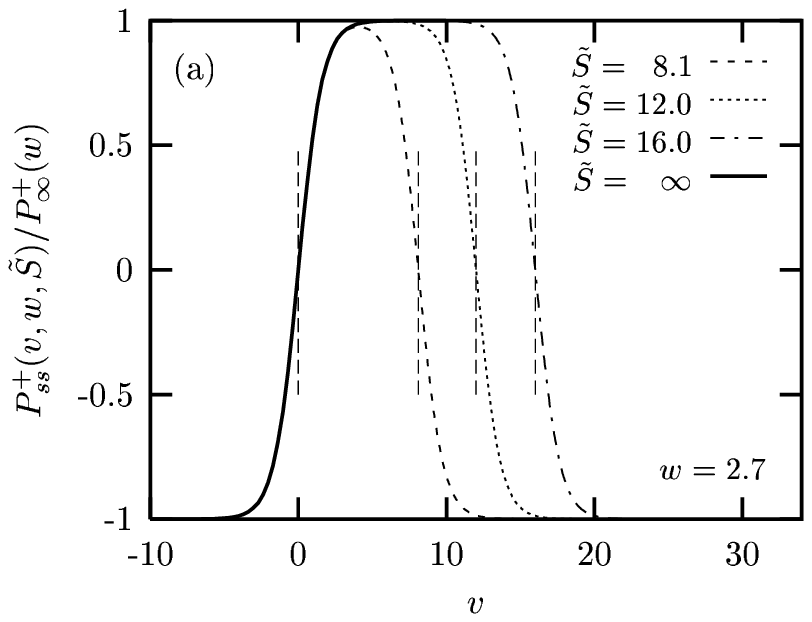,width=0.5\textwidth} \quad
      \epsfig{file=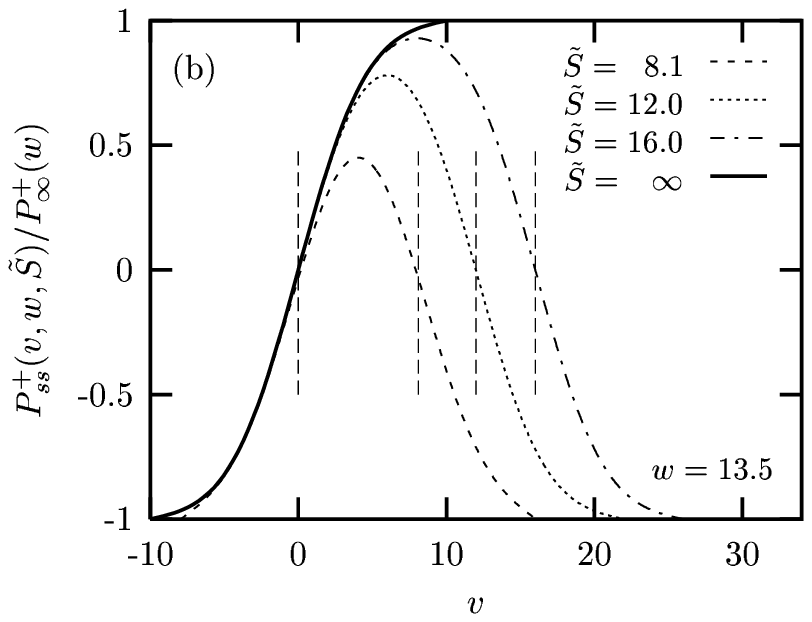,width=0.5\textwidth}
      \caption{\label{fig:step-stripe}Comparison between the normalized cross sections of the order 
               parameter scaling function near a substrate with negative surface field in which a stripe with
               positive surface field of scaled width $\tilde{S}$ is embedded and the normalized cross 
               sections of the order parameter profile near a single chemical step (which emerges as
               limiting case for $\tilde{S} \to \infty$), (a) at a scaled normal distance $w=2.7$, (b) 
               at $w=13.5$. Here we consider the case $T > T_c$. The chemical steps are located at $x=0$
               and $x=S$ corresponding to $v=0$ and $v=\tilde{S}$ (see vertical lines).}
    \end{center}
  \end{figure}

%
The second system we focus on consists of a laterally extended substrate with a negative surface field 
$h_1 \to -\infty$ in which a stripe of width $S$ with a positive surface field 
$h_1 \to \infty$ is embedded (see Fig.\,\ref{fig:geometry}(b)).
The surface field $h_1$ is assumed to vary step-like:
  \begin{eqnarray}
     h_1(x) = \left \{
              \begin{array}{rl}
                + \infty \quad, & x \in [0,S] \\[1mm]
                - \infty \quad, & x \notin [0,S] \,.
              \end{array}
              \right.
     \label{eq:h1-stripe-scaled} 
  \end{eqnarray}
We introduce the coordinates $v$ and $w$ scaled in units of the correlation length, where $v$ denotes 
the scaled distance from the left border of the stripe at $x=0$ and $w$ the scaled distance from the 
substrate. The influence of the stripe width $S$ and of the reduced temperature $t$ are captured by 
the scaled stripe width $\tilde{S}= S/\xi^\pm$. The order parameter profile for the system with a single 
stripe (ss) exhibits the scaling property
  \begin{eqnarray}
    && \phi(x,z,t,S) \nonumber \\
    && = a |t|^\beta P^\pm_{ss}\bigg(v=\frac{x}{\xi^\pm}, w=\frac{z}{\xi^\pm}, \tilde{S}=\frac{S}{\xi^\pm}\bigg)
    \label{eq:scalingbehavoir-2d-stripe}
  \end{eqnarray}
with the limiting behavior
  \begin{eqnarray}
     P^\pm_{ss}(v, w, \tilde{S}=0) &=& - P^\pm_\infty (w) \,, \\[1mm]
     P^\pm_{ss}(v, w, \tilde{S} \to \infty) &=& P^\pm_{cs}(v,w) \,.
  \end{eqnarray}
Figures\,\ref{fig:step-stripe}(a) and \ref{fig:step-stripe}(b) show cross sections (parallel to the 
substrate) through the order parameter scaling function $P^+_{ss}(v,w,\tilde{S})$ for systems with different 
scaled stripe width $\tilde{S}$ at two different scaled distances $w$ from the substrate in comparison 
with a cross section through the order parameter profile $P^+_{cs}(v,w)$ at a single chemical step 
at $v=0$ (see Subsec.\,\ref{subsec:chem_step}).
With increasing scaled width $\tilde{S}\to \infty$ of the stripe the left part of the cross 
section of the stripe system merges with the cross section of the system with a single step. Further away 
from the substrate the mutual influence of the two step structures onto each other is more pronounced 
(see Fig.\,\ref{fig:step-stripe}(b)) and stronger for smaller stripes.

For systems with a single chemical stripe on the substrate a reference system with the scaling function 
$P^{+}_{ss,0}(v,w,\tilde{S})$ can be introduced similar to the one for a single chemical step which can be 
interpreted as a system with a single chemical stripe on the substrate where no healing occurs:
  \begin{eqnarray} 
    P^{+}_{ss,0}(v,w,\tilde{S}) = \left\{  
      \begin{array}{ll}
        + P^+_\infty(w), & |v| \in [0,\tilde{S}] \\[2mm]
        - P^+_\infty(w), & |v| \notin [0,\tilde{S}] \,.
      \end{array} \right.
    \label{eq:P_ss0}
  \end{eqnarray}
This reference system allows us to define the excess adsorption $\tilde{\Gamma}^+_{ex,ss}$ for the 
striped systems:
   \begin{eqnarray}
     \tilde{\Gamma}^+_{ex,ss} (\tilde{S})
     = \int_{-\infty}^\infty \! \dd v \int_{0}^\infty \! \dd w \!\!\! 
       && \left( P^+_{ss}(v,w,\tilde{S}) \right. \nonumber \\
       && \left. - P^+_{ss,0}(v,w,\tilde{S}) \right) \,.
     \label{eq:gammass}
   \end{eqnarray}
  \begin{figure}
    \begin{center}
      \epsfig{file=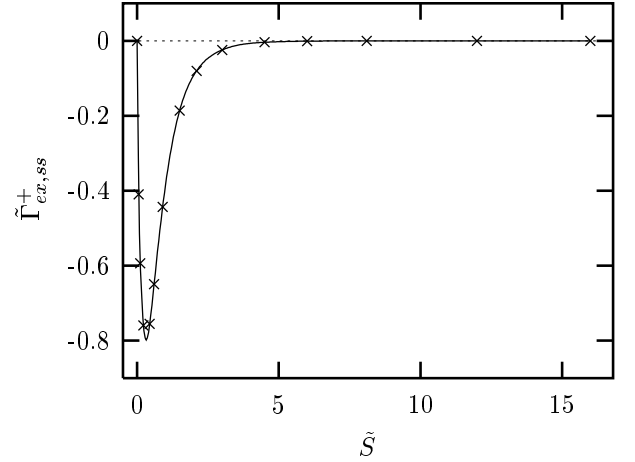,width=0.5\textwidth}\\
      \caption{\label{fig:excess_stripe-stepped}Universal excess adsorption $\tilde{\Gamma}^+_{ex,ss}$  
               (Eq.\,(\ref{eq:gammass})) for a single stripe with respect to a system with no healing
               as a function of the stripe width $\tilde{S}$. For $\tilde{S}=0$ the system corresponds 
               to a system with a homogeneous substrate whose corresponding excess adsorption is 0. In 
               the limit $\tilde{S} \to \infty$ the stripe system corresponds to two independent chemical 
               steps whose excess adsorption also vanishes due to antisymmetry of the profiles around 
               $v=0$ and $v=\tilde{S} \to \infty$. The non-vanishing excess adsorption for 
               intermediate stripe widths $\tilde{S}$ indicates the effect of the stripe with a 
               surface field opposite to the one of the embedding substrate. $\tilde{\Gamma}^+_{ex,ss}$
               attains its minimum at $\tilde{S} \simeq 0.3$.}
    \end{center}
  \end{figure}
Equation (\ref{eq:gammass}) can be rewritten as
   \begin{eqnarray}
     \tilde{\Gamma}^+_{ex,ss} (\tilde{S})
     = \lim_{\tilde{B} \to \infty} \left\{ 
         \int_{-\tilde{B}/2}^{\tilde{B}/2} \! \dd v \int_{0}^\infty \! \dd w \, P^+_{ss}(v,w,\tilde{S}) \right. \nonumber \\
         \left. + (\tilde{B} - 2\tilde{S}) \tilde{\Gamma}^+_\infty \,,
       \right\} \,
     \label{eq:gammass_2}
   \end{eqnarray}
where $B=\tilde{B}\xi \gg S$ is the overall lateral extension of the substrate surface in $x$-direction
(see Fig.\,\ref{fig:geometry}(b)) and $\tilde{\Gamma}^+_\infty = \int_{0}^\infty \dd w \, P^+_\infty(w)$ 
is a universal number characterizing the amplitude of the excess adsorption at a \textit{homogenous} 
substrate: $\int_0^\infty \dd z \, \phi(z,t) = \tilde{\Gamma}^+_\infty a \xi^+_0 \, t^{\beta-\nu}$. 
The value of $\tilde{\Gamma}^+_\infty$ is discussed in Ref. \cite{Smith1997} and in Ref. \cite{Floeter1995} 
where it is denoted as $\tilde{\Gamma}^+_{\infty} = g_+ / (\nu-\beta)$ (see Eq. (2.9) and Fig.\,5 in 
Ref. \cite{Floeter1995}; in $d=3$ one has $\tilde{\Gamma}^+_{\infty} = 2.27$). Equation (\ref{eq:gammass_2}) 
describes how in an operational sense the universal function $\tilde{\Gamma}^+_{ex,ss} (\tilde{S})$ 
(see Fig.\,\ref{fig:excess_stripe-stepped}) can be obtained from the measurements of the excess 
adsorption (relative to the bulk order parameter) at a striped surface and from those of the 
excess adsorption at the corresponding homogeneous surface.

\begin{figure}
    \begin{center}
      \epsfig{file=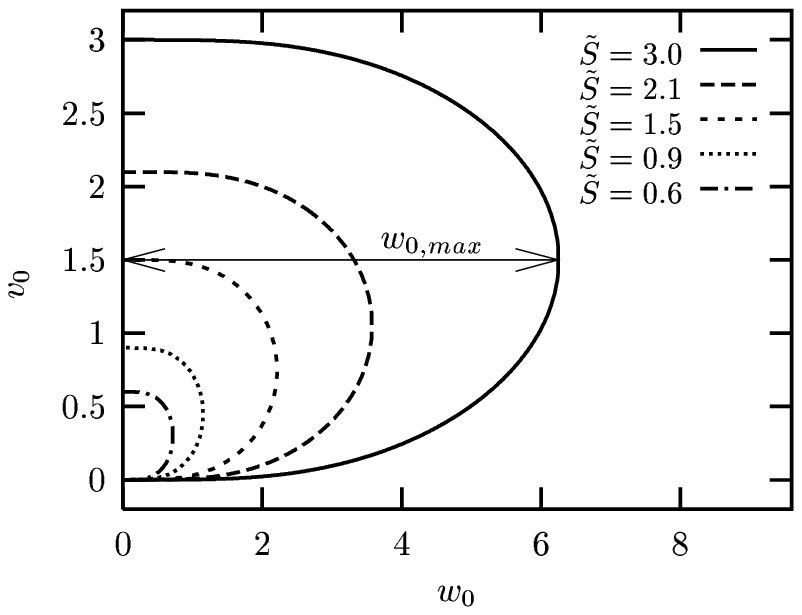,width=0.5\textwidth}\\
      \caption{\label{fig:tongue}Zero-lines $v_0(w)$ where the order parameter vanishes: $P(v_0,w)=0$
               for different stripe widths $\tilde{S}$. Outside the tongue the order parameter is 
               negative corresponding to the preference of the substrate outside the stripe. Inside the 
               tongue the order parameter is positive and thus demarcates the range of the influence of 
               the stripe with opposite preference.}
      \epsfig{file=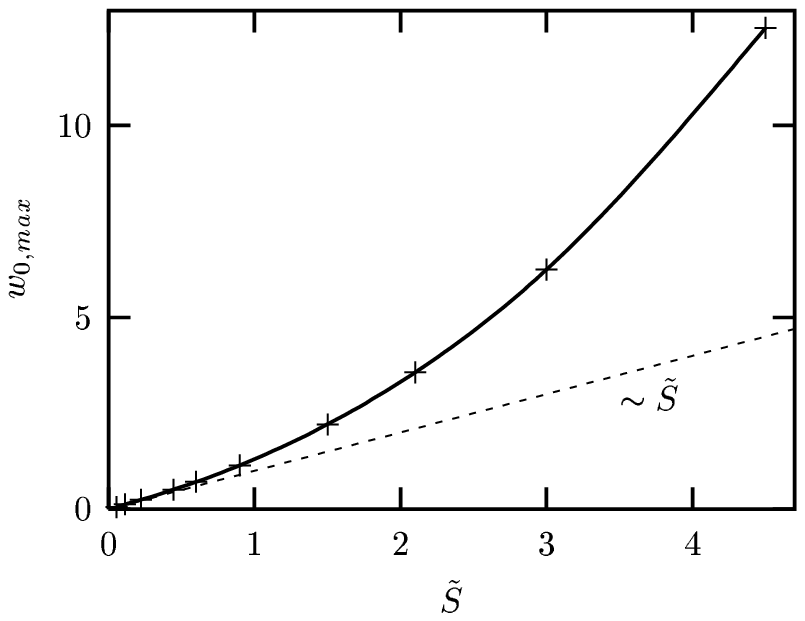,width=0.5\textwidth}\\
      \caption{\label{fig:tongue-length}Dependence of the length $w_{0,max}$ of the tongues shown in 
               Fig.\,\ref{fig:tongue} on the stripe width $\tilde{S}$. For small stripe widths the length 
               of the tongue increases linearly,
               with increasing stripe widths $\tilde{S}$ the length $w_{0,max}$ of the tongues diverges.}
    \end{center}
  \end{figure}

Figure\,\ref{fig:excess_stripe-stepped} shows the dependence of the excess adsorption 
$\tilde{\Gamma}^+_{ex,ss}$ on the scaled stripe width $\tilde{S}$. For $\tilde{S}\to \infty$
the structures associated with the two chemical steps forming the stripe decouple so that
$\tilde{\Gamma}_{ex,ss}(\tilde{S}\to \infty)=0$. Upon construction $\tilde{\Gamma}_{ex,ss}(\tilde{S}=0)=0$.
The decrease of the excess adsorption at $\tilde {S} \approx 1$  arises due to the fact that for 
sufficiently small stripes the spatial region where the order parameter profile is positive does no 
longer resemble a rectangular but a tongue-like shape (see Fig.\,\ref{fig:tongue}). Since for the reference 
system the region with a positive order-parameter resembles still a rectangle this leads to a negative 
excess adsorption.

For different stripe width $\tilde{S}$ the tongue-like regions defined by the zero-lines $v_0(w)$ 
where the order parameter vanishes, $P(v_0,w)=0$, are shown in Fig.\,\ref{fig:tongue}.
The width of the tongue at the substrate is given by the stripe width $\tilde{S}$ due to the 
infinitely strong surface fields. With increasing stripe width $\tilde{S}$ the tongue becomes longer.
Fig.\,\ref{fig:tongue-length} shows the dependence of the length $w_{0,max}$ of the tongue on the stripe 
width $\tilde{S}$. It shows that for small stripes $w_{0,max}$ increases linearly with $\tilde{S}$.

\subsection{Periodic stripes}
%
As a natural extension we now consider a substrate with a periodic array of stripes with alternating
surface fields (see Fig.\,\ref{fig:geometry}(c)): stripes of width $S_p$ ($S_n$) with positive 
(negative) surface field $h_1 \to \infty$ ($h_1 \to -\infty$). In the lateral $x$-direction we employ periodic boundary 
conditions. The  corresponding scaling function $P_{ps}$ for the order parameter distribution near
the substrate with periodic stripes (ps) depends on two scaled coordinates $v=x/\xi^+$ and $w=z/\xi^+$ and 
on two scaled stripe widths $\tilde{S}_p=S_p/\xi^+$ and $\tilde{S}_n= S_n/\xi^+$ or, equivalently, 
$\tilde{S}_p$ and $\tilde{S}_n/\tilde{S}_p=S_n/S_p$. Thus in the series $P_\infty$, $P_{cs}$, $P_{ss}$, and $P_{ps}$ each 
scaling function acquires one additional scaling variable.
Considering the excess adsorption $\tilde{\Gamma}^+_{ex,ps}(\tilde{S}_p, \tilde{S}_n/\tilde{S}_p)$ reduces the number 
of scaling variables to two:
   \begin{eqnarray}
     \tilde{\Gamma}^+_{ex,ps} && \hspace{-6mm} (\tilde{S}_p, \tilde{S}_n/\tilde{S}_p)  \nonumber \\[1mm]
     &=& N \int_0^{\tilde{S}_p+\tilde{S}_n} \dd v \int_{0}^\infty \dd w   
           \left\{ P^+_{ps}(v, w, \tilde{S}_p, \tilde{S}_n/\tilde{S}_p) \right. \nonumber \\[1mm]
     & & \quad \left. - P^+_{ps,0}(v, w, \tilde{S}_p, \tilde{S}_n/\tilde{S}_p) \right\} \!,
     \label{eq:gammaps}
   \end{eqnarray}
where $P^+_{ps,0}=P^+_\infty(w)$ on the positive stripe and $P^+_{ps,0} = -P^+_\infty(w)$  on the 
negative stripe. $N$ is the number of periodic cells on the substrate. In analogy to Eq.\,(\ref{eq:gammass_2}) 
one has
  \begin{eqnarray}
     \tilde{\Gamma}^+_{ex,ps} && \hspace{-6mm}(\tilde{S}_p, \tilde{S}_n/\tilde{S}_p)  \nonumber \\[1mm]
     &=& N \left\{ \int_0^{\tilde{S}_p+\tilde{S}_n} \dd v \int_{0}^\infty \dd w \, 
            P^+_{ps}(v, w, \tilde{S}_p, \tilde{S}_n/\tilde{S}_p) \right. \nonumber \\[1mm]
     & & \quad \left. - (\tilde{S}_p-\tilde{S}_n) \tilde{\Gamma}^+_\infty \right\} \,.
     \label{eq:gammaps_2}
   \end{eqnarray}

  \begin{figure}
    \begin{center}
      \epsfig{file=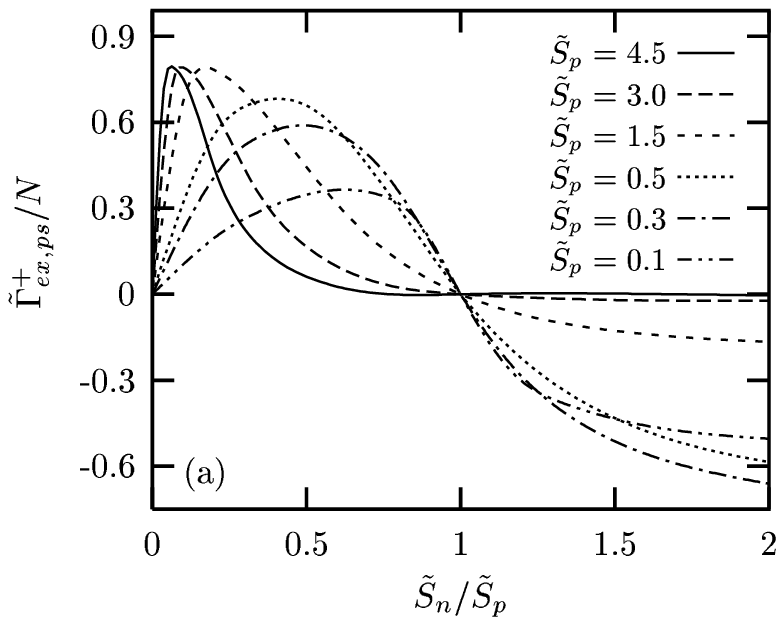,width=0.5\textwidth}
      \epsfig{file=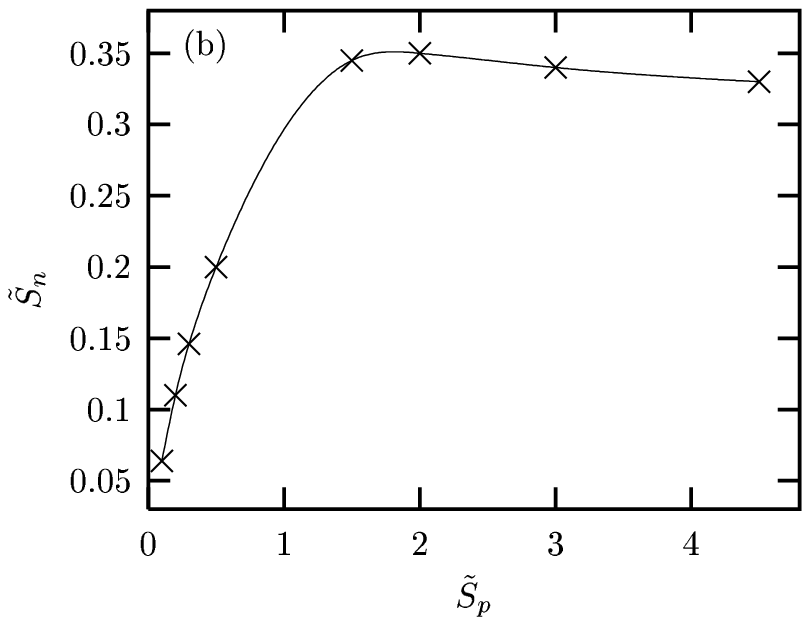,width=0.5\textwidth}
      \caption{\label{fig:peri_stripe-stepped}(a) Universal scaling function for the excess adsorption 
               $\tilde{\Gamma}^+_{ex,ps}$ per unit cell (Eqs.\,(\ref{eq:gammaps}) and (\ref{eq:gammaps_2})) 
               for a system with a periodic stripe pattern of negative and positive surface fields with 
               scaled widths $\tilde{S}_n=S_n/\xi$ and $\tilde{S}_p=S_p/\xi$, respectively. The case 
               $\tilde{S}_n/\tilde{S}_p = 0$ corresponds to a homogeneous substrate with a positive 
               surface field. The excess adsorption vanishes for $\tilde{S}_n=\tilde{S}_p$ due to 
               symmetry reasons. For large ratios $\tilde{S}_n/\tilde{S}_p$ with $\tilde{S}_p$ fixed 
               the excess adsorption $\tilde{\Gamma}^+_{ex,ps}/N$ tends to the excess adsorption of a 
               single stripe of width $\tilde{S}_p$ which is given by Fig.\,\ref{fig:excess_stripe-stepped}.
               (b) Loci of the maxima of $\tilde{\Gamma}^+_{ex,ps}(\tilde{S}_p, \tilde{S}_n/\tilde{S}_p)/N$.}
    \end{center}
  \end{figure}

As compared with the case of a single stripe the periodic arrangement enhances the excess adsorption
by the (potentially large) number of lateral repeat units. Figure\,\ref{fig:peri_stripe-stepped}(a) shows
that $\tilde{\Gamma}^+_{ex,ps} \gtrless 0$ for $S_p \gtrless S_n$. $\tilde{\Gamma}^+_{ex,ps}$ vanishes at
$S_n=0$ because this corresponds to the limiting case of a homogeneous substrate. $\tilde{\Gamma}^+_{ex,ps}$
also vanishes for $S_n=S_p$ due to symmetry reasons. The limit $\tilde{S}_n/\tilde{S}_p \to \infty$
with $\tilde{S}_p$ fixed leads to the case of a single stripe of width $S_p$; this corresponds to 
Fig.\,\ref{fig:excess_stripe-stepped} up to the factor $N$. In Fig.\,\ref{fig:peri_stripe-stepped}(b) the
loci of the maxima of the excess adsorption $\tilde{\Gamma}^+_{ex,ps}/N$ for different stripe widths $\tilde{S}_p$
and $\tilde{S}_n$ are given.

\begin{figure}
    \begin{center}
      \epsfig{file=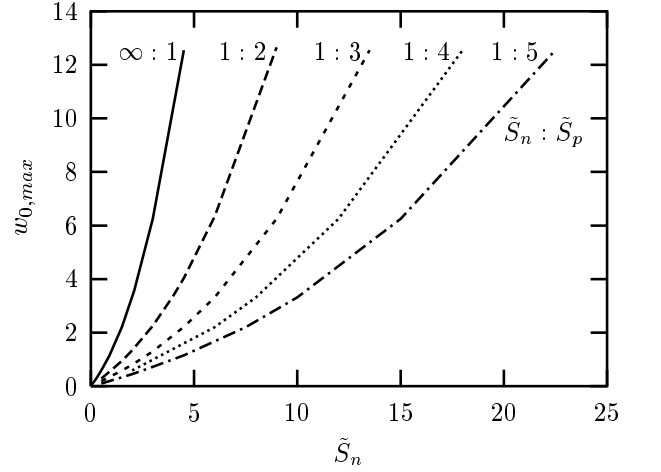,width=0.5\textwidth}
      \caption{\label{fig:peri_stripe-tongue}Dependence of the rescaled length $w_{0,max}$ of the 
               tongues within which the order parameter adjacent to a stripe maintains the sign preference 
               of the stripe (compare Fig.\,\ref{fig:tongue}) on the stripe width $\tilde{S}_n$ for 
               different ratios $\tilde{S}_n/\tilde{S}_p$. The limit $\tilde{S}_n/\tilde{S}_p \to \infty$ 
               corresponds to the case of a single stripe of width $\tilde{S}_p$. The full curve corresponds to the 
               latter case (see also Fig.\,\ref{fig:tongue-length}) and is denoted as $\infty:1$.}
    \end{center}
  \end{figure}

The positive values of the excess adsorption $\tilde{\Gamma}^+_{ex,ps}$ are again caused by the
tongue-like regions within which the order parameter profile has a definite sign. For 
$\tilde{S}_p < \tilde{S}_n$ there is a finite tongue-like region adjacent to each positive stripe within which
the order parameter is positive and negative outside of it. For $\tilde{S}_p \to \tilde{S}_n$ the tongue
length diverges and the tongue boundaries degenerate into two parallel lines orthogonal to the substrate.
For $\tilde{S}_p > \tilde{S}_n$ there are tongues of negative values of the order parameter adjacent to the 
negative stripes.
Figure \ref{fig:peri_stripe-tongue} shows the length $w_{0,max}$ of the tongues on the stripe width
$\tilde{S}_n$ for different ratios $\tilde{S}_n/\tilde{S}_p$. With increasing ratio $\tilde{S}_n/\tilde{S}_p \to \infty$
the limit of a substrate with a single positive stripe of width $\tilde{S}_p$ in a negative matrix is reached.

\section{\label{sec:summary}Summary}
%
Based on mean field theory combined with renormalization group arguments we have studied critical 
adsorption of fluids at chemically structured substrates. The fluids are either one- or two-component
liquids near their gas-liquid critical point or binary liquid mixtures near their critical demixing 
point. In the first case the order parameter is given by the local total density, in the second case 
by the local concentration. We have determined the order parameter profiles and suitably defined
excess adsorptions for three substrate types: a single chemical step, a single chemical stripe,
and a periodic stripe pattern (see Fig.\,\ref{fig:geometry}). We have obtained the following main 
results:
\begin{enumerate}
\item
The order parameter profiles and the excess adsorption can be described in terms of universal scaling
functions (Eqs.\,(\ref{eq:scalingbehavior-2d}), (\ref{eq:gamma_ex-2}), (\ref{eq:gammacs_gammabulk}) - 
(\ref{eq:gammacs_minus}), (\ref{eq:scalingbehavoir-2d-stripe}), (\ref{eq:gammass_2}),
(\ref{eq:gammaps_2})). The excess adsorptions are introduced relative to order parameter profiles
at homogeneous substrates (Sec.\ref{sec:hom-sub} and Fig.\,\ref{fig:half-space}) taken to vary 
step-like in lateral direction according to the actual chemical pattern under consideration.
\item
The specific shapes of the scaling functions are determined within mean field theory. For the 
\textit{chemical step} the full scaling function of the order parameter profiles is shown in 
Fig.\,\ref{fig:3d-plot} in terms of the scaling variables $v=x/\xi$ and $w=z/\xi$ given by the 
lateral ($x$) and the normal ($z$) coordinates in units of the bulk correlation length $\xi$. 
Lateral cuts through the normalized scaling function with an emphasis on its asymptotic behavior
far from the substrate are shown in Fig.\,\ref{fig:cross}. For the case of strong adsorption
considered here the slopes of the scaling function across the chemical step increase $\sim w^{-2}$ 
upon approaching the surface and decay $\sim e^{-Cw}$ towards the bulk with $C=1$ above $T_c$ and $C<1$ 
below $T_c$ (Fig.\,\ref{fig:slope}). To a large extent the variation of the full scaling function 
normal to the surface can be absorbed by rescaling the lateral variation suitably 
(Fig.\,\ref{fig:slope-norm}). The excess adsorption at the chemical step (Eq.\,(\ref{eq:integ-ex}) 
and Fig.\,\ref{fig:integration}) lead to universal numbers above (Eq.\,(\ref{eq:gammacs_plus})) and 
below $T_c$ (Eqs.\,(\ref{eq:gammacs_gammabulk}) and (\ref{eq:gammacs_minus})).
\item
The lateral variation of the order parameter adjacent to a \textit{single chemical stripe} is shown in
Fig.\,\ref{fig:step-stripe} in terms of its suitably normalized scaling function. 
Figure\,\ref{fig:step-stripe} visualizes the dependence of these structures on the scaled stripe width.
Figure\,\ref{fig:tongue} illustrates the influence of a chemical stripe of width $S$ on the adjacent order 
parameter. The range of this influence, defined as the spatial region of maintaining the preferred sign of
the order parameter, generates tongue-like structures which grow with increasing stripe width 
(Fig.\,\ref{fig:tongue-length}). The excess adsorption (Eq.\,(\ref{eq:gammass_2})) is described by a 
universal scaling function in terms of $\tilde {S}=S /\xi$ which is maximal for $\tilde{S}\simeq 0.3$ 
(Fig.\,\ref{fig:excess_stripe-stepped}).
\item
For a \textit{periodic stripe pattern} of $N$ unit cells the scaling function for the order parameter 
depends on four scaling variables: $v$, $w$, $\tilde{S}_p=S_p/\xi$, and $\tilde{S}_n=S_n/\xi$ where 
$S_p$ and $S_n$ are the width of the stripes with positive and negative surface fields, respectively. 
The range of influence (see 3. above) of the narrower stripes is again confined to tongue-like structures
which grow with increasing stripe width (Fig.\,\ref{fig:peri_stripe-tongue}). The corresponding 
excess adsorption (Fig.\,\ref{fig:peri_stripe-stepped}(a)) is given by a universal scaling function in 
terms of $\tilde{S}_p$ and $\tilde{S}_n$ which describes the interpolation between the homogeneous substrate 
($\tilde{S}_n/\tilde{S}_p=0$) and a single stripe of width $\tilde{S}_p$ ($\tilde{S}_n/\tilde{S}_p=\infty$, 
$\tilde{S}_p$ fixed). The relation between $\tilde{S}_p$ and $\tilde{S}_n$ which yields the maximum excess
adsorption is shown in Fig.\,\ref{fig:peri_stripe-stepped}(b).
\end{enumerate}

\appendix
\section{\label{app}Numerical methods}
%
In the following, we provide some details of the numerical methods which we have applied. The order 
parameter profiles are be calculated numerically from Eqs.\,(\ref{eq:diff-eq}) - (\ref{eq:bound-cond-bulk}) 
by introducing a discrete lattice with finite spatial extensions. The extension of the system perpendicular 
to the substrate ($z$-direction) is $L$, the extension in the direction of the inhomogeneity of the 
substrate ($x$-direction) is $B$. The corresponding lattice spacings are denoted as $dz$ and $dx$. 
Since the system is translationally invariant in the $d-2$ directions perpendicular to the $x-z$-plane 
(where $d$ denotes the spatial dimension of the system), the numerical problem is effectively 
two-dimensional. \\
\indent For the calculation of the scaling functions these quantities are scaled with the correlation length 
$\xi^\pm$ leading to the scaled length $\tilde{L}$ in $w$-direction, the scaled width $\tilde{B}$ in 
$v$-direction, and the corresponding lattice spacings $dw$ and $dv$, respectively.
In order to mimic the characteristics of an infinitely extended system we choose an exponentially 
decaying continuation of the order parameter profiles as the boundary condition at the distance $L$ 
from the substrate. (For the non-antisymmetric profiles we resort to a constant continuation.) 
In the case of the substrate with a single chemical step (cs) or a single stripe (ss) we choose the 
width $\tilde{B}$ such that the system is sufficiently broad so that the influence
of the chemical steps at the lateral boundaries is negligible and the order parameter attains the 
value of the corresponding half-space profile $P^\pm_\infty(w)$ for infinite surface fields 
$h_1 \to \infty$ or $P^\pm_{h_1}(w)$ for finite surface fields $h_1 < \infty$:
$P^\pm_{cs}(\pm \frac{\tilde{B}}{2},w) = \pm P^\pm_{\infty,h_1}(w)$, 
$P^\pm_{ss}(\pm \frac{\tilde{B}}{2},w) = - P^\pm_{\infty,h_1}(w)$.
For the substrate with periodic stripes one can focus on a single unit cell so that the scaled width 
$\tilde{B}$ is the sum of the stripe widths $\tilde{S}_p$ and $\tilde{S}_n$ with periodic boundary 
conditions: $P^\pm_{ps}(+\frac{\tilde{B}}{2},w) = P^\pm_{ps}(-\frac{\tilde{B}}{2},w)$.
The choices for the width $\tilde{B}$ and the length $\tilde{L}$ of the system are not completely 
independent, i.e., for a given length $\tilde{L}$ there is a minimum width $\tilde{B}$ so that  
$\tilde{L}$ and $\tilde{B}$ span a region in which the order parameter profile is calculated correctly 
under the chosen boundary conditions. It turns out that a width $\tilde{B} \gtrsim 16$ is sufficient 
for the studied range of lengths $\tilde{L} \lesssim 20$. For the substrates with a single stripe the 
width $\tilde{B}-\tilde{S}$ of the negative matrix is chosen as $\tilde{B}-\tilde{S} \gtrsim 16$ 
and is kept constant for different widths of the stripe.
We use the steepest descent method in order to calculate the order parameter profiles. The values of 
the scaling function $P(v,w)$  of the order parameter at each lattice point are split into an initial 
part $P_{in}(v,w)$ (see the profiles for systems where no healing occurs given by Eqs.\,(\ref{eq:P_cs0})
and (\ref{eq:P_ss0})), which is the known solution for a system similar to the one under consideration, 
and a correction term $P_{corr}(v,w)$ which is varied in accordance with the steepest descent method. 
This procedure is described in detail in Ref. \cite{Schlesener2003}.
The excess adsorption depends on the value of the lattice spacings $\dd v$ and $\dd w$, respectively, 
used for its numerical calculation. Therefore we have calculated the excess adsorption for different 
lattice constants and extrapolated it to $\dd v=\dd w=0$. It also depends on the length of the system. 
Hence we calculated the order parameter profiles for a fixed width $\tilde{B}$ and different lengths 
$\tilde{L}$. For lengths larger than $\tilde{L}=10$ the results are indistinguishable, i.e., for these 
lengths corrections with respect to an infinitely long system are negligible.





\begin{thebibliography}{99}
\bibitem{liquid-vapor}
This terminology refers to the case of a simple fluid below the critical temperature where within mean 
field theory the corresponding scaling function for the interface between the liquid phase and the vapor 
phase is given by Eq.\,(\ref{eq:lv-interface}).
\end{thebibliography}
\end{document}